\documentclass[aps,prd,twocolumn,showpacs,floatfix,preprintnumbers,amsmath,
amssymb]{revtex4}

\usepackage{amsmath,amssymb}
\usepackage{epsfig}

\def\bea{\begin{eqnarray}}
\def\eea{\end{eqnarray}}

 \newcommand{\CI}{\mathbb{C}}

\newcommand{\PI}{\mathbb{P}}

\newcommand{\hphi}{\hat{\phi}}
\def\etal{{\it et al}.}  
   
\def\goesas{\mathop{\sim}\limits}  
\def\Z#1{_{\lower2pt\hbox{$\scriptstyle#1$}}}
\def\X#1{_{\lower2pt\hbox{$\scriptscriptstyle#1$}}}

\begin{document}

\preprint{gr-qc/0506103}

\title{Thermodynamics and Stability of Higher Dimensional \\
Rotating (Kerr) AdS Black Holes}
\author{Benedict M.N. Carter and Ishwaree P. Neupane \\}

\affiliation{ Department of Physics and Astronomy, University of
Canterbury, Private Bag 4800, Christchurch, New Zealand}

\date{July 7, 2005}

\begin{abstract}

We study the thermodynamic and gravitational stability of Kerr
anti-de Sitter black holes in five and higher dimensions. We show,
in the case of equal rotation parameters, $a_i=a$, that the
Kerr-AdS background metrics become stable, both thermodynamically
and gravitationally, when the rotation parameters $a_i$ take
values comparable to the AdS curvature radius. In turn, a Kerr-AdS
black hole can be in thermal equilibrium with the thermal
radiation around it only when the rotation parameters become not
significantly smaller than the AdS curvature radius. We also find
with equal rotation parameters that a Kerr-AdS black hole is
thermodynamically favored against the existence of a thermal AdS
space, while the opposite behavior is observed in the case of a
single non-zero rotation parameter. The five dimensional case is
however different and also special in that there is no high
temperature thermal AdS phase regardless of the choice of rotation
parameters. We also verify that at fixed entropy, the temperature
of a rotating black hole is always bounded above by that of a
non-rotating black hole, in four and five dimensions, but not in
six and more dimensions (especially, when the entropy approaches
zero or the minimum of entropy does not correspond to the minimum
of temperature). In this last context, the six dimensional case is
marginal.

\end{abstract}

\pacs{98.80.Cq}

\maketitle

\section{Introduction}
Black holes are perhaps the most tantalizing objects in general
relativity. Recently, the study of black holes in a background
anti-de Sitter spacetime has been well motivated from developments
in string/M-theory, which naturally incorporate black holes as
solitonic D-branes, or simply branes as the higher-dimensional
progenitors of black holes. An intriguing example of this is the
conjectured duality~\cite{Maldacena97a} between string theory on
$AdS_5\times S^5$ background and ${\cal N}=2$ Super Yang-Mills
theory in four dimensions, and in particular, Witten's
interpretation~\cite{Witten98} of the Hawking-Page phase
transition between thermal AdS and AdS black
hole~\cite{Hawking83a} as the confinement-deconfinement phases of
the dual gauge theory defined on the asymptotic boundaries of the
AdS space.

Much effort has been put into the weak AdS gravity regime,
analyzing the implications of AdS black holes on dual (gauge)
theories at non-zero temperature, using the conjectured AdS/CFT
correspondence. In this context, the most interesting black hole
solutions are presumably the five dimensional Kerr-AdS solutions
for a stationary black hole~\cite{Hawking98a}. The thermodynamics
of AdS quantum gravity has been extensively used to infer the
thermodynamics of quantum field theory in the large $N$ (or weak
field) limits, with an AdS gravity dual, such as
Schwarzschild-AdS~\cite{Witten98},
Kerr-Newman-AdS~\cite{Hawking:1999dp,Mann:1999,
Awad:1999xx,Landsteiner:1999xv} and
hyperbolic-AdS~\cite{Danny98a,Cai:2001jc,Ish03a} black holes.

The Kerr metric~\cite{Kerr63a} is a simple explicit exact solution
of the Einstein vacuum equations describing a rotating black hole
in a four-dimensional flat spacetime. Shortly after Kerr's
discovery, Carter~\cite{Carter68a} provided an elegant
generalization of the Kerr solution in four-dimensional de-Sitter
and anti-de Sitter backgrounds. A higher dimensional
generalization of Kerr metric in a flat background was given by
Myers and Perry~\cite{Myers86a}. But its generalization to five
and higher dimensions with a non-zero cosmological constant was
given, only recently, by Hawking {\it et al.} ~\cite{Hawking98a}
and Gibbons {\it
  et al.}~\cite{Gibbons04a,Gibbons04b}. There has been recent
  interest in constructing the analogous charged rotating
  solutions in gauged supergravity in
four, five and seven dimensions~\cite{Cvetic05a}, and also on
non-uniqueness~\cite{Ross04a} of those solutions in five and
higher dimensions.

The study of non-charged rotating (Kerr) black holes is
interesting at least for two reasons. Firstly, the thermodynamics
of Kerr black holes, in a background AdS space, can give rise to
interesting descriptions in terms of CFTs defined on the
(conformal) boundary of AdS space, leading to a better
understanding of the AdS/CFT correspondence~\cite{Hawking98a}.
Secondly, astronomically relevant black hole spacetimes are, to a
very good approximation, may be described by the Kerr metric.

As not much is known about the stability of Kerr-AdS black holes
in higher dimensions, in this paper we study the thermodynamic
stability of these black holes in five and higher dimensions. We
also investigate the gravitational stability of a background
Kerr-(A)dS spacetime under metric perturbations.

The layout of the paper is as follows. We begin in Sec. II by
outlining the (anti)-de Sitter background metrics in $d$
dimensions and their generalizations to Kerr-AdS solutions. In
Sec. III we pay special attention to the thermodynamic stability
of Kerr-AdS black holes by studying the behavior of Hawking
temperature, free energy and specific heat in various dimensions.
In Section IV we study the gravitational stability of background
Kerr-(A)dS metrics under linear tensor perturbations. The
stability of a rotating anti-de Sitter background spacetime in
dimensions higher than four was not previously studied. Our
linearized perturbation equations have other interesting
applications. In particular, they allow us to study the stability
of background AdS metrics with non-trivial rotation parameters.

Separability of Hamilton-Jacobi and Klein-Gordon equations in the
Kerr (anti)-de Sitter backgrounds was discussed
in~\cite{Vasudevan:2004ca}, especially in the limit when all
rotation parameters take the same value, see~\cite{Kunduri:2005fq}
for a discussion in five dimensions. An earlier work on
separability of the Hamilton-Jacobi equation and quantum radiation
from a five dimensional Kerr black hole with two rotation
parameters, but in an asymptotically flat background, can be found
in~\cite{Frolov02a}. However, our analysis is different. It
corresponds not to a separability of the wave equations for a
particle but rather to a separability of radial and angular wave
equations under linear tensor perturbations.

\section{AdS and Kerr-AdS metrics}

One of the interesting features of the Kerr metric in (anti)-de
Sitter spaces is that it can be written in the so-called
Kerr-Schild form, where the metric $g_{a b}$ is given exactly by
its linear approximation around the (anti)-de Sitter metric
$\tilde{g}_{a b}$ as follows~\cite{Gibbons04a,Vasudevan:2004ca}:
\begin{equation} \label{Kerr-Schild1}
 ds^2= g_{ab}\, dx^a dx^b =
\tilde{g}_{a b} {dx}^a {dx}^b + \frac{2M}{U}
\left(k_a dx^a \right)^2,
\end{equation}
where $k_a$ is a null geodesic with respect to both the full
metric $g_{a b}$ and the (A)dS metric $\tilde{g}_{a b}$. Moreover,
the Ricci tensor of $g_{a b}$ is related to that of $\tilde{g}_{a
b}$ by
\begin{eqnarray}
R_a\,^b&=&\tilde{R}_a\,^b - \tilde{R}_a\,^c  h_c\,^b \nonumber \\
&{}& +\frac{1}{2}\left(\tilde{\nabla}_c \tilde{\nabla}_a h^{b c} +
\tilde{\nabla}^c \tilde{\nabla}^b h_{a c}-
\tilde{\nabla}^c\tilde{\nabla}_c h_a\,^b\right),
\end{eqnarray}
where $h_{ab}= \frac{2M}{U} k_a k_b$, with $M$ and $U$ being the
parameters proportional to the mass and gravitational potential of
a Kerr black hole respectively. Thus, the stability of a Kerr
metric under metric perturbation is specific to the stability of
the background metric, which is given by $M=0$.

Let us begin with a five dimensional (anti)-de Sitter metric in
the standard form:
\begin{eqnarray}\label{5dAdS}
\widetilde{d{\rm s}}^2&=& -(1+c y^2 )dt^2
+\frac{{dy}^2}{1+cy^2}\nonumber \\
&{}&+\,y^2\left(\frac{dx^2}{1-x^2}+ (1-x^2){d{\hat\phi_1}}^2+ x^2
{d{\hat\phi_2}}^2\right),
\end{eqnarray}
which satisfies $R_{\mu\nu}=-4 c g_{\mu\nu}$, with $c>0$ in AdS
space. The apparent singularities at $x=\pm 1$ are merely
coordinate singularities. By defining $x= \cos\hat\theta$, one
sees that the coordinate $x$ has range $-1\leq x\leq 1$ while
$(\hat\phi_1, \hat\phi_2)$ have a period $2\pi$, so
$(x,\hat\phi_1)$ parameterizes (topologically) a $2$-sphere, while
$\hat\phi_2$ parameterizes an $S^1$-fiber.

The metric (\ref{5dAdS}) is easily generalized to six and higher
dimensions. In six dimensions, one has
\begin{equation}
\widetilde{d{\rm s}}^2= -(1+cy^2)dt^2+\frac{dy^2} {1+cy^2}+y^2 d\Sigma_4^2,
\end{equation} where
\begin{equation}
d\Sigma_4^2= \frac{d\Omega^2}{1-x^2}+ (1-x^2) d\hphi_3^2 +
\hphi_1^2 d\psi^2,
\end{equation}
\begin{equation}
d\Omega^2 = (1-\hphi_2^2) d\hphi_1^2
+(1-\hphi_1^2) d\hphi_2^2+ 2
\hphi_1 \hphi_2\,d\hphi_1
d\hphi_2,
\end{equation}
and $x^2=\hphi_1^2+\hphi_2^2$. The apparent singularities at
$x=\pm 1$ are again merely coordinate singularities. By defining
$$\hphi_1= \sin\theta \sin\varphi,
\quad \hphi_2= \sin\theta\cos\varphi, $$
one easily sees that $(\hphi_1,\hphi_2,\psi)$ parameterize
(topologically) a $3$-sphere. In fact, the
generalized (anti)-de Sitter metric can be written
in a more compact form:
\begin{equation}\label{compact1}
\widetilde{ds}^2= -(1+c y^2 )dt^2
+\frac{{dy}^2}{1+cy^2}+y^2\sum_{k=1}^{N+\varepsilon}
\left(d\hat{\mu}_k^2+\hat{\mu}_k^2 d\hat{\phi}_k^2\right)
\end{equation}
satisfying
\begin{equation}
\sum_{i=1}^{N+\varepsilon} \hat{\mu}_i^2=1,
\end{equation}
where $N=(d-1)/2$, $\varepsilon=0$ (if $d$ is odd), or
$N=(d-2)/2$, $\varepsilon=+1$ (if $d$ is even). Both in odd and
even dimensions, there are $N$ azimuthal coordinates
$\hat{\phi}_i$, each with period $2\pi$. But when $d$ is even
there is an extra coordinate $\hat{\mu}_{N+1}$, which lies in the
interval $-1 \le \hat{\mu}_{N+1} \le 1$.

In $AdS_d$ spaces the rotation group is $SO(d-1)$ and the number
of independent rotation parameters for a localized object is equal
to the number of Casimir operators, which is the integer part of
$(d-1)/2$. Thus in four dimensions the metric of a Kerr black hole
can have only one Casimir invariant of the rotation group $SO(3)$,
which is uniquely defined by an axis of rotation, while in five
dimensions it can have two independent rotation parameters
associated with two possible planes of rotation.

One may introduce to~(\ref{compact1}) $N$ rotation parameters, for
example, using the following coordinate transformation:
\begin{equation} \label{first-co-trans}
y^2= \sum_{i=1}^N \frac{(r^2+a_i^2)\mu_i^2}{1-c a_i^2},
\end{equation}
where $\sum_{i=1}^{N+\varepsilon} \mu_i^2=1$. The constants $a_i$
which are introduced in~(\ref{first-co-trans}) merely as
parameters in a coordinate transformation may be interpreted as
genuine rotation parameters after one adds to~(\ref{compact1}) the
square of an appropriate null vector, as in~(\ref{Kerr-Schild1}).
Using the following coordinate transformations~\cite{Hawking98a}:
\begin{eqnarray}
&&dt=d\tau +\frac{2M}{V-2M}\frac{dr}{(1+cr^2)}, \nonumber \\
&&d\hat{\phi_i}= d\phi_i+ca_i d\tau +\frac{2M}{V-2M} \frac{a_i
dr}{(r^2+a_i^2)},
\end{eqnarray}
and combining the expressions~(\ref{Kerr-Schild1}),
(\ref{compact1}), (\ref{first-co-trans}), one would obtain the
Kerr (A)dS metrics in Boyer-Lindquist coordinates. We are not
going into details of this construction but refer to
Ref.~\cite{Gibbons04a} for an elegant discussion. In five
dimensions, the metric of Kerr-AdS solution is
\begin{eqnarray}\label{transform1}
&&ds^2=-W(1+cr^2) d\tau^2 +\frac{\rho^2 dr^2}{V-2M}
+\frac{\rho^2}{\Delta_\theta} d\theta^2\nonumber \\
&& ~~~~~+\sum_{i=1}^2\frac{r^2+a_i^2}{1-ca_i^2}\,
\mu_i^2(d\phi_i+ca_i d\tau)^2\nonumber \\
&& ~~~~~+\frac{2M}{\rho^2}\left( d\tau - \sum_{i=1}^{2} \frac{a_i
\mu_i^2 d\phi_i}{1-ca_i^2}\right)^2,
\end{eqnarray}
where $\mu_1=\cos\theta$, $\mu_2=\sin\theta$,
\begin{eqnarray}
&&\rho^2=r^2+a_1^2 \cos^2\theta+a_2^2\sin^2\theta, \nonumber \\
&& \Delta_{\theta}=1- ca_1^2\cos^2\theta -ca_2^2 \sin\theta^2,
\nonumber \\
&& V= \frac{1}{r^2} (1+cr^2)(r^2+a_1^2)(r^2+a_2^2),\nonumber \\
&& W= \frac{\Delta_{\theta}}{\Xi_1\Xi_2},\quad \Xi_i=1-c a_i^2.
\end{eqnarray}
In the limit $a_i\to 0$, one recovers the standard
Schwarzschild-AdS metric. As we see shortly, black holes with
non-zero rotation parameters, or, in general, Kerr-AdS black
holes, enjoy many interesting properties distinct from
Schwarzschild-AdS black holes.

\section{Thermodynamics of Kerr AdS Solutions}

Using the standard technique of background subtraction, Gibbons
\etal~\cite{Gibbons04c} have recently calculated the regularized
(Euclidean) actions for the Kerr-AdS black holes in arbitrary
$d~(\geq 4)$ dimensions. The results are
\begin{equation}
\hat{I}=-\frac{{\cal A}_{d-2}}{8\pi{G} \prod_j\Xi_j
}\frac{\beta}{l^2}\left({l}^{2N} \prod_{i=1}^{N}(R^2+\alpha_i^2)-m
l^2  \right)
\end{equation}
for odd $d (=2N+1)$, and
\begin{equation}
\hat{I} =-\frac{{\cal A}_{d-2}}{8\pi{G} \prod_j\Xi_j
}\frac{\beta}{l} \left( R\, {l}^{2N} \prod_{i=1}^{N}
(R^2+\alpha_i^2)-m l \right)
\end{equation}
for even $d (=2N+2)$, where
\begin{equation}
{\cal A}_{d-2}=\frac{2\pi^{(d-1)/2}}{\Gamma[(d-1)/2]},
\end{equation}
is the volume of the unit $(d-2)$-sphere. In the above we have
defined $c\equiv 1/l^2$, with $l$ being the curvature radius of
the (bulk) AdS space. The dimensionless parameters are:
$\Xi_j\equiv 1-\alpha_j^2, ~R\equiv r_+/l$ and $\alpha_i\equiv
a_i/l$., where, as usual, $r_+$ is the radius of the horizon,
which occurs at a root of $V-2M=0$, and $m\equiv M(r=r_+)$. The
Hawking temperature, which is the inverse of Euclidean period,
$T\equiv 1/\beta$, is given by
\begin{equation}
T= \frac{R}{2\pi l} \left[(1+R^2)\left(\sum_{i=1}^{N}
\frac{1}{R^2+\alpha_i^2}+\frac{\varepsilon}{2 R^2}\right)
-\frac{1}{R^2}\right],
\end{equation}
where $\varepsilon=0$ for odd $d$ and $+1$ for even $d$.

\begin{figure}[ht]
\epsfig{figure=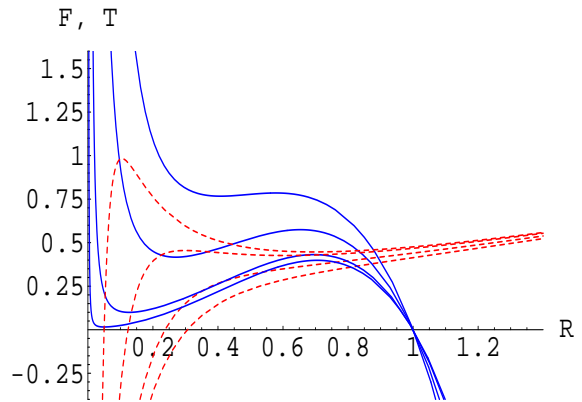,height=2.3in,width=3.0in}
\caption{($d=5$) The free energy (solid lines) and temperature
(dashed lines) as a function of horizon position, with
$\alpha_1=\alpha_2\equiv \alpha$. From top to bottom (free energy)
or bottom to top (temperature): $\alpha=1/3,\,0.25,\,1/8,\,0.05$.
In all plots we have set $4G=1$.} \label{figure1}
\end{figure}
\begin{figure}[ht]
\epsfig{figure=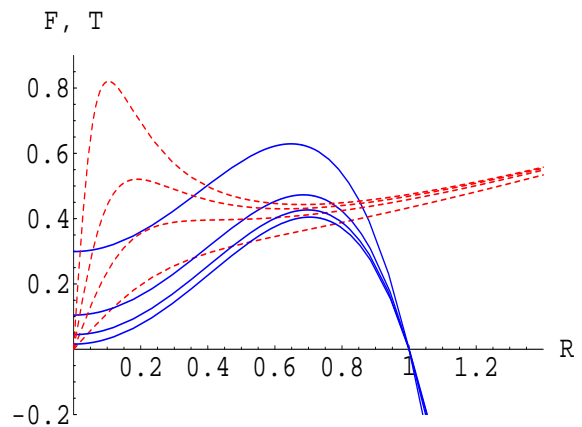,height=2.3in,width=3.0in}
\caption{($d=5$) The free energy (solid lines) and temperature
(dashed lines) vs horizon position, with $\alpha_1\equiv \alpha$,
$\alpha_2=0$. From top to bottom (free energy) or bottom to top
(temperature): $\alpha=0.4,\,0.25,\,1/6,\,0.1$.} \label{figure2}
\end{figure}

\begin{figure}[ht]
\epsfig{figure=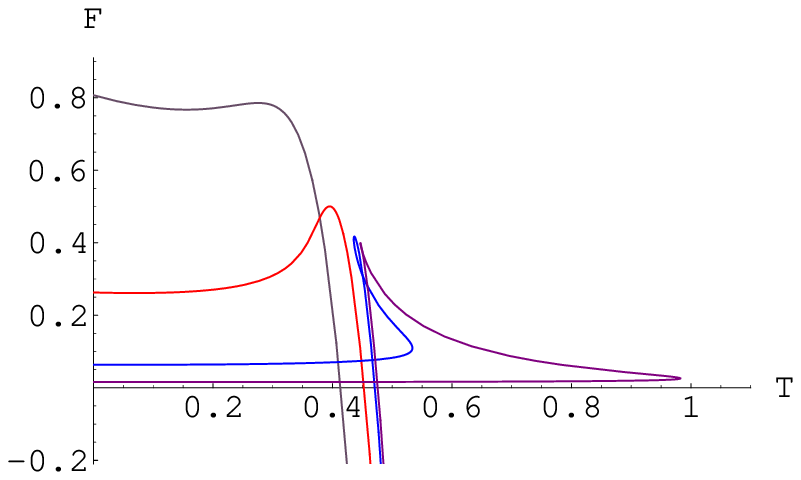,height=2.3in,width=3.0in}
\caption{($d=5$) The free energy vs temperature
with $\alpha_1=\alpha_2\equiv
\alpha $. From top to bottom: $\alpha=1/3,\,0.2,\,0.1,\,0.05$}
\label{figure3}
\end{figure}

\begin{figure}[ht]
\epsfig{figure=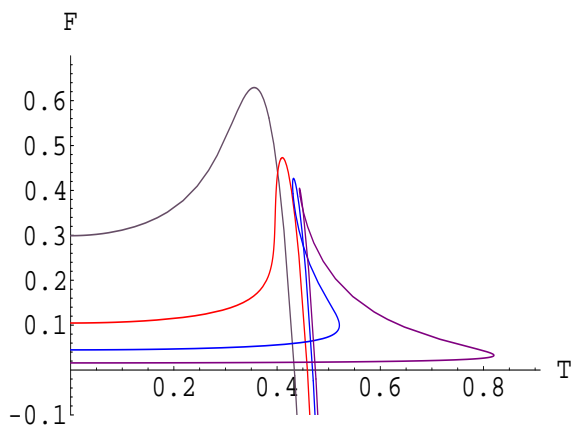,height=2.3in,width=3.0in}
\caption{($d=5$) The free energy vs temperature with
$\alpha_1\equiv \alpha $, $\alpha_2=0$. From top to bottom:
$\alpha=0.4,\,0.25,\,1/6,\,0.1$} \label{figure4}
\end{figure}

\medskip

The calculation of total energy in an asymptotically (A)dS
background is somewhat trickier (see e.g.~\cite{Gibbons04c}),
mainly because the analogous Komar integral for the relevant
time-like Killing vector diverges, which then requires a
regularization, see also Ref.\cite{Papa:2005} which presents a
general analysis for the conserved charges and the first law of
thermodynamics for the four dimensional Kerr-Newman-AdS and the
five dimensional Kerr-AdS black holes. In this context, the
conserved charges (energies) $E$ and $E^\prime$ associated with
different Killing vectors, respectively, $\partial_t$ and
$\partial_t+l^{-1} \alpha_i
\partial_{\phi_i}$ are different. However, the
calculation of free energy itself is unambiguous. In fact, one can
always identify the free energy of a Kerr-AdS black hole as
$F=\hat{I}\times 1/\beta$, and hence
\begin{equation}\label{free-energy-gen}
F=\frac{{\cal A}_{d-2}}{16\pi{G}} \frac{(l\,R)^{d-3}
\left(1-R^2\right)} {\prod_{j}\Xi_j} \prod_{i=1}^{N}
\left(1+\frac{\alpha_i^2}{R^2}\right).
\end{equation}
This result is modified from that of a Schwarzschild-AdS black
hole by certain terms in the product which are now functions of
$R$ and the rotation parameters $\alpha_i$.

\subsection{Thermal Phase Transition}

In four spacetime dimensions black holes are stable (see
e.g.~\cite{Gregory93a}), but the issue of stability may be raised
in five and higher dimensions. The five-dimensional Kerr-AdS
solutions are particularly interesting as these could be embedded
into IIB supergravity in ten dimensions.

\begin{figure}[ht]
\epsfig{figure=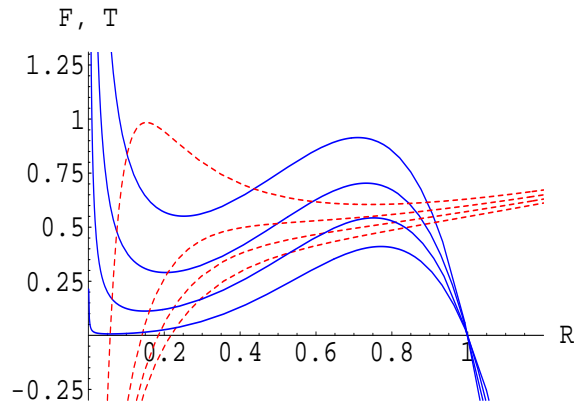,height=2.3in,width=3.0in}
\caption{($d=6$) The free energy and temperature vs horizon
position, with $\alpha_1=\alpha_2=\alpha$. From top to bottom:
$\alpha=0.4,\,1/3,\,0.25,\,0.1$.} \label{figure5}
\end{figure}

\begin{figure}[ht]
\epsfig{figure=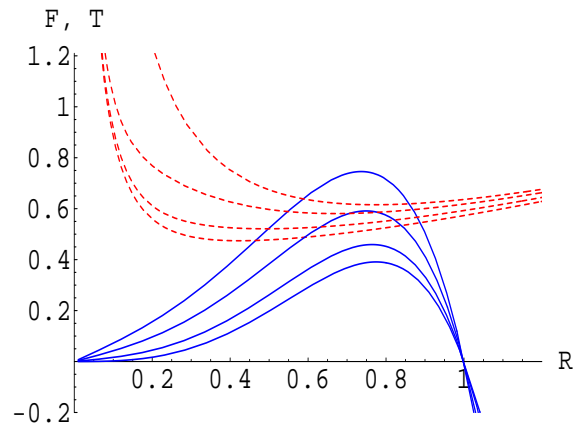,height=2.3in,width=3.0in}
\caption{($d=6$) The free energy and temperature vs horizon
position, with $\alpha_1\equiv \alpha$, $\alpha_2=0$. From top to
bottom: $\alpha=0.5,\,0.4,\,0.25,\,0.04$.} \label{figure6}
\end{figure}

\begin{figure}[ht]
\epsfig{figure=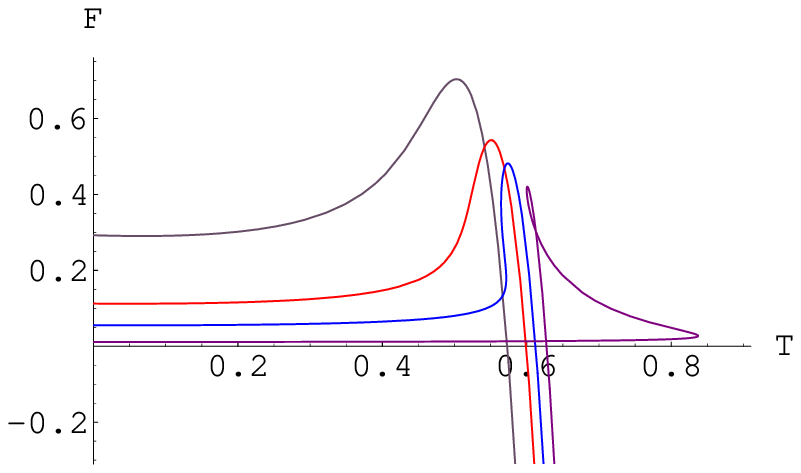,height=2.3in,width=3.0in}
\caption{($d=6$) The free energy vs temperature with
$\alpha_1=\alpha_2\equiv \alpha $. From top to bottom:
$\alpha=1/3,\,0.25,\,0.2,\,0.12$.} \label{figure7}
\end{figure}

\begin{figure}[ht]
\epsfig{figure=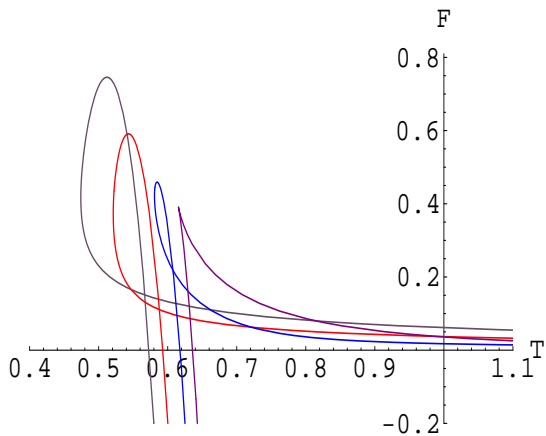,height=2.3in,width=3.0in}
\caption{($d=6$) The free energy vs temperature with
$\alpha_1=\alpha $, $\alpha_2=0$. From left to right:
$\alpha=0.5,\,0.35,\,0.2,\,0.05$} \label{figure8}
\end{figure}

From~(\ref{free-energy-gen}), it is readily seen that a phase
transition between the background AdS space and the black hole is
set by the scale $R=1$, so that $R>1$ corresponds to AdS black
hole ($F<0$) and $R<1$ to a thermal AdS space ($F>0$). This
behavior may be seen also in terms of the charge or potential if
present. In general, when the values of the rotation parameters
$\alpha_i$ are decreased, the free energy lowers towards zero at
low temperature. For $0<\alpha\ll1$, in the small $R$ region, $F$
nearly approaches but never touches the $F=0$ axis (see Figs. $1$
and $2$). That is, the free energy curve crosses the $F=0$ axis
only once, namely when $R=1$, which usually corresponds to the
Hawking-Page phase transition point. But, in dimensions $d\geq 6$,
this alone does not mean that a first order phase transition of
Hawking-Page type is essentially present.

In five dimensions, with $\alpha_1=\alpha_2\equiv \alpha >0$,
there is a minimum $R$ below which the temperature appears to be
negative and also diverges as $R\to 0$ (see Fig.~\ref{figure1}),
which is clearly unphysical. In fact, there is a minimum in
temperature below which the Kerr black holes simply do not exit.
Nevertheless, the plots in Fig.~\ref{figure3} show the free energy
can be a well defined function of temperature. We also note that
the specific heat is a monotonically increasing function of
temperature when $\alpha\geq 0.17 $.

The Hawking temperature of a Kerr-AdS black hole with a
non-vanishing rotation parameter approaches zero as $R$ goes to
zero. The free energy is still a smooth function of both the
horizon size and the temperature (see Figs.~\ref{figure2} and
\ref{figure4}). These all imply a thermodynamic stability of a
small Kerr black hole in $AdS_5$ space.

\begin{figure}[ht]
\epsfig{figure=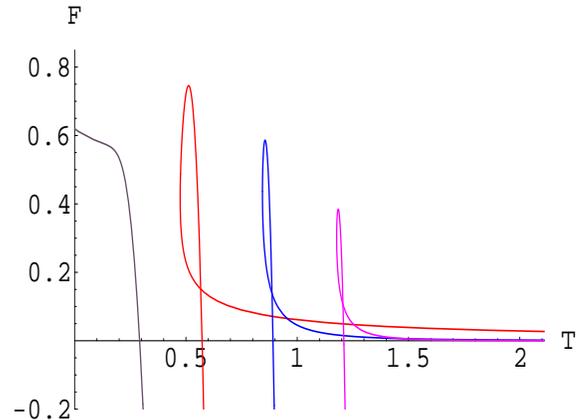,height=2.3in,width=3.0in} \caption{The
free energy vs temperature with a single non-vanishing rotation
parameter $\alpha$. From left to right $d=4$ (with $\alpha=0.3$)
and $d=6,\,8,\,10$ (with $\alpha=0.5$).} \label{figure9}
\end{figure}

\begin{figure}[ht]
\epsfig{figure=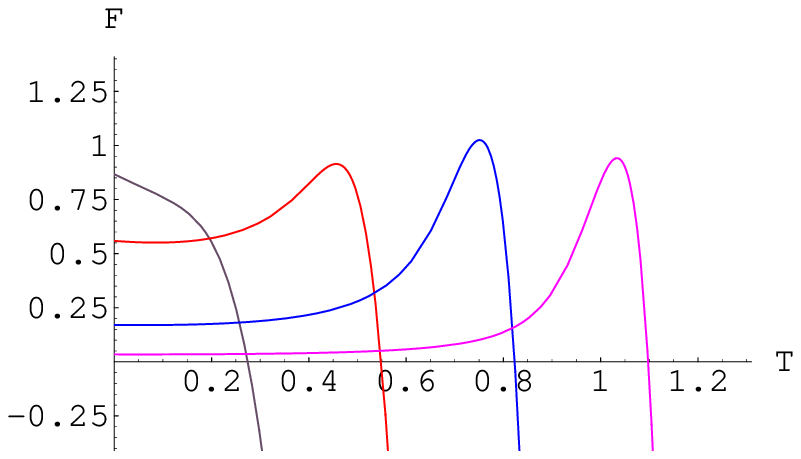,height=2.3in,width=3.0in} \caption{The
free energy vs temperature with equal rotation parameters,
$\alpha_i=\alpha=0.4$. From left to right $d=4,\,6,\,8,\,10$.}
\label{figure10}
\end{figure}

\begin{figure}[ht]
\epsfig{figure=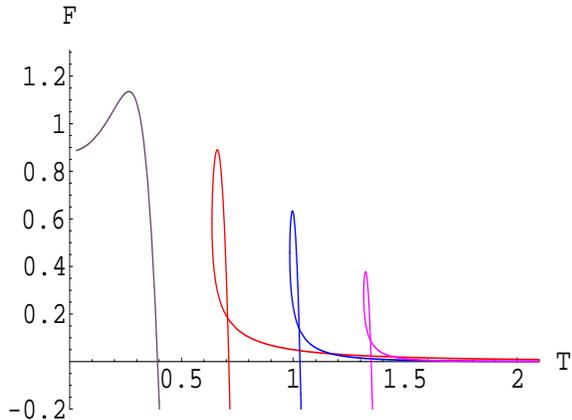,height=2.3in,width=3.0in} \caption{The
free energy vs temperature with a single non-vanishing rotation
parameter $\alpha$. From left to right: $d=5,\,7,\,9,\,11$ (with
$\alpha=0.6$).} \label{figure11}
\end{figure}

\begin{figure}[ht]
\epsfig{figure=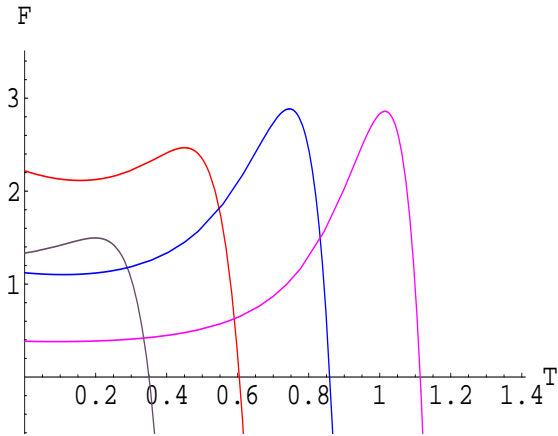,height=2.3in,width=3.0in} \caption{The
free energy vs temperature with equal rotation parameters,
$\alpha_i=\alpha=0.5$. From left to right $d=5,\,7,\,9,\,11$.}
\label{figure12}
\end{figure}

The thermodynamic behavior above is essentially the opposite in
six dimensions, where the temperature always diverges at $R= 0$.
As the plots in Figs.~\ref{figure5}--\ref{figure8} show the
thermodynamics of single parameter solutions are quite different
from those with equal rotation parameters. We also note that the
$F<0$ region in Figs.~\ref{figure9}--\ref{figure12} corresponds to
$R>1$, while the region $F\geq 0$ corresponds to $0\leq R\leq 1$.
When $d\geq 6$, in the case of equal rotation parameters, only
small black holes are globally preferred and locally stable, while
in the case of single rotation parameter, a thermal AdS phase is
more preferred. The behavior in five dimensions is special in that
there is no high temperature thermal AdS phase regardless of the
choice of rotation parameters.

\subsection{The first law of AdS bulk thermodynamic}

One of the simplest ways of calculating the energy in an
asymptotically AdS background is to integrate the first law of
(bulk) thermodynamics:
\begin{equation}\label{firstlaw}
dE=T~dS+\sum_i \Omega_i J_j.
\end{equation}
where the entropy $S$ and angular momenta (of a rotating black
hole) $J_i$ are defined via
\begin{equation}
S= \beta \frac{\partial\hat{I}}{\partial\beta}-\hat{I}, \quad
J_i=-\frac{\partial F}{\partial\Omega_i},
\end{equation}
where
\begin{equation}\label{ang-velo}
\Omega_i\equiv \frac{\alpha_i (1+R^2) l}{R^2+\alpha_i^2}.
\end{equation}
In Ref.~\cite{Gibbons04c}, the mass (energy) of Kerr AdS black
hole was evaluated, by demanding {\it as a priori} that entropy of
the black hole is one-quarter the area, $S=A/4$, in order to
satisfy~(\ref{firstlaw}). The results are
\begin{eqnarray}\label{Energy-odd-d}
&& d=2N+1=\mbox{odd}: \nonumber \\
&&E=\frac{m {\cal A}_{d-2}}{4\pi \Pi_j\Xi_j} \left(\sum_{i=1}^N
\frac{1}{\Xi_i}-\frac{1}{2}\right),\nonumber \\
&& S=\frac{{\cal A}_{d-2}}{4} \left(l\,R\right)^{2N-1}
\prod_{i=1}^{N}\left(1+\frac{\alpha_i^2}{R^2}\right)\frac{1}{\Xi_i},
\end{eqnarray}
\begin{eqnarray}\label{Energy-even-d}
&& d= 2N+2=\mbox{even}: \nonumber \\
&& E=\frac{m {\cal A}_{d-2}}{4\pi \Pi_j\Xi_j} \sum_{i=1}^{N}
\frac{1}{\Xi_i},\nonumber \\
&& S=\frac{{\cal A}_{d-2}}{4} \left(l\,R\right)^{2N}
 \prod_{i=1}^{N}\left(1+\frac{\alpha_i^2}{R^2}\right)\frac{1}{\Xi_i}.
\end{eqnarray}
This result differs from the expression of energy suggested by
Hawking {\it et al.} in~\cite{Hawking98a}, both in odd and even
dimensions,
\begin{equation}\label{Energy-Hawking}
E^\prime=\frac{m {\cal A}_{d-2}}{4\pi \prod_{j=1}^{N}
\Xi_{j}}\frac{(d-2)}{2}.
\end{equation}
The reason for this is that the energy~(\ref{Energy-Hawking}) is
measured in a frame rotating at infinity with the angular
velocities:
\begin{equation}\label{ang-velo-Hawking}
\Omega_i^\prime= \frac{\alpha_i\Xi_i \,l}{R^2+\alpha_i^2},
\end{equation}
instead of~(\ref{ang-velo}). Since the angular velocities differ
by $\Omega_i-\Omega_i^\prime=\alpha_i \,l$, the two results above,
(\ref{Energy-odd-d}) or (\ref{Energy-even-d}),
(\ref{Energy-Hawking}), agree only in the limit $\alpha_i\to 0$
(i.e. $\Sigma_i\to 1$).

A remark is in order. The energy of background AdS spacetime (i.e.
$m=0$) is expected to be the same as the Casimir energy of a dual
field theory in one dimension lower, up to a conformal factor. But
from the above result one finds $E=0$ when $m=0$. To understand
this apparent discrepancy, it should be noted that the ADM mass
$M$ is only a local definition of black hole energy, while the
total energy of a localized object in a curved background normally
takes into account the asymptotic value of the background itself
(which is non-zero in the AdS space). And, in general, one can
write $E=M+E_0$, where $E_0$ is an integration constant. For
example, for a Schwarzschild-AdS black hole with hyperbolic
symmetry ($k=-1$), $E_0$ may be given by $E_0=-M_{e}$, where
$M_{e}$ is the black hole mass at the extremal limit, see
e.g.~\cite{Ish03a}.

\begin{figure}[ht]
\epsfig{figure=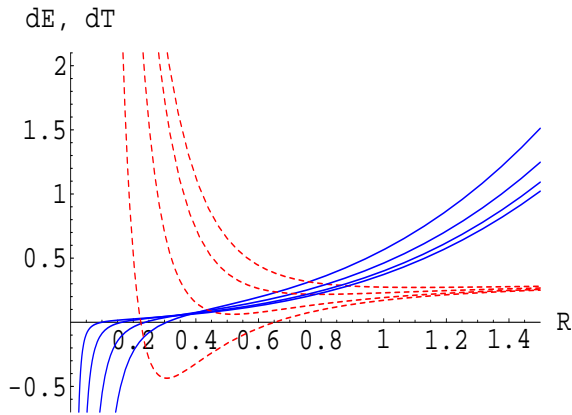,height=2.3in,width=3.0in}
\caption{($d=5$) The energy and temperature differentials vs
horizon position, with $\alpha_1=\alpha_2\equiv \alpha $. As $R\to
0$, $dE\to +\infty$ and $dT\to -\infty$. From left to right:
$\alpha=0.1,\,1/6,\,0.25,\,1/3$.} \label{figure13}
\end{figure}

\begin{figure}[ht]
\epsfig{figure=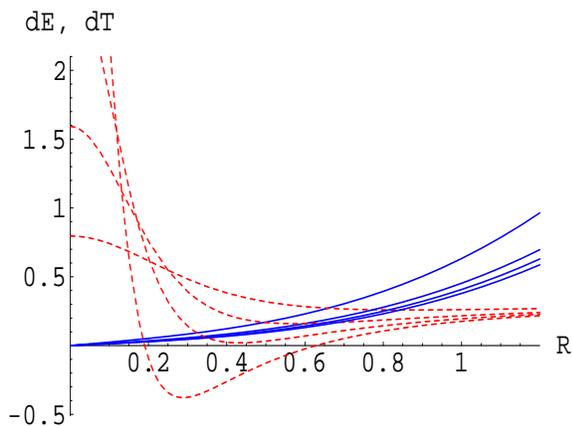,height=2.3in,width=3.0in}
\caption{($d=5$) The energy and temperature differentials vs
horizon position, with $\alpha_1=\alpha $, $\alpha_2=0$. As $R\to
0$, $dE\to 0$ and $dT>0$. From top to bottom (free energy) or
bottom to top (temperature) $\alpha=0.5,\,1/3,\,0.25,\,1/6$.}
\label{figure14}
\end{figure}

\begin{figure}[ht]
\epsfig{figure=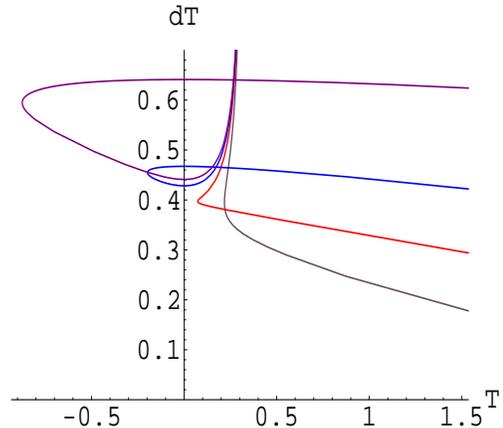,height=2.3in,width=3.0in}
\caption{($d=5$) The temperature differential vs temperature with
$\alpha_1= \alpha_2=\alpha$; from top to bottom
$\alpha=0.08,\,0.12,\, 0.17,\,0.25$} \label{figure15}
\end{figure}

\begin{figure}[ht]
\epsfig{figure=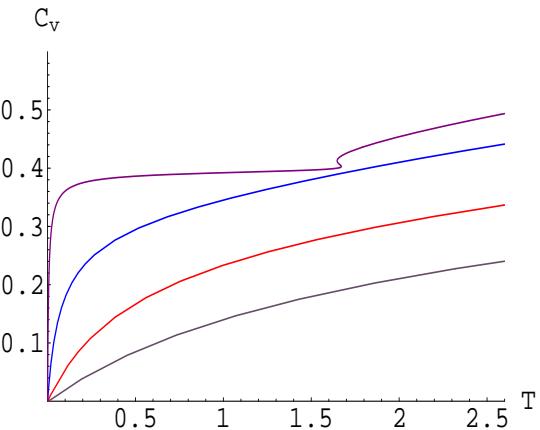,height=2.3in,width=3.0in}
\caption{($d=5$) The specific heat vs temperature with
$\alpha_1=\alpha_2=\alpha$; from top to bottom
$\alpha=0.17,\,1/3,\,0.5,\,0.6$. When $\alpha$ is $\lesssim 1/6$,
then there would appear a new branch with almost constant specific
heat at low temperature. } \label{figure16}
\end{figure}

\begin{figure}[ht]
\epsfig{figure=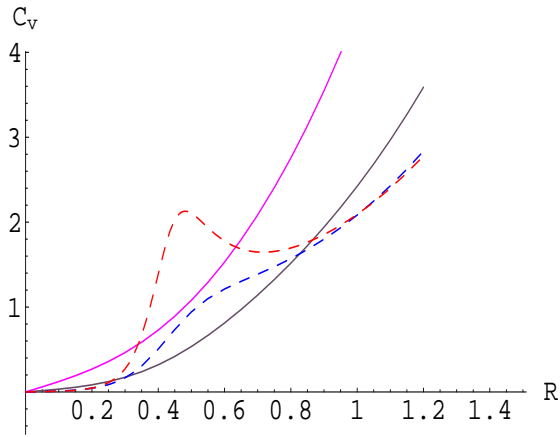,height=2.3in,width=3.0in}
\caption{($d=5$) The specific heat vs horizon position with
$\alpha_1=\alpha$, $\alpha_2=0$. From top to bottom:
$\alpha=0.7,\,0.5$ (solid lines); and bottom to top: $0.3,\,0.26$
(dashed lines). With $\alpha\gtrsim 0.245$, the specific heat
curve has a single branch and it is a monotonically increasing
function of Hawking temperature.} \label{figure17}
\end{figure}

\begin{figure}[ht]
\epsfig{figure=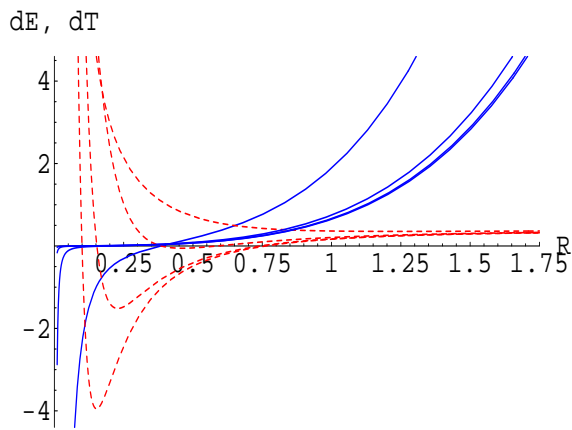,height=2.3in,width=3.0in}
\caption{($d=6$) The energy differential (solid lines) and
temperature differential (dashed lines) vs horizon position, with
$\alpha_1=\alpha_2\equiv \alpha $. From left to right ($dE$ in the
$dE>0$ region) or top to bottom ($dT$):
$\alpha=0.5,\,0.2,\,0.1,\,1/15$} \label{figure18}
\end{figure}

\begin{figure}[ht]
\epsfig{figure=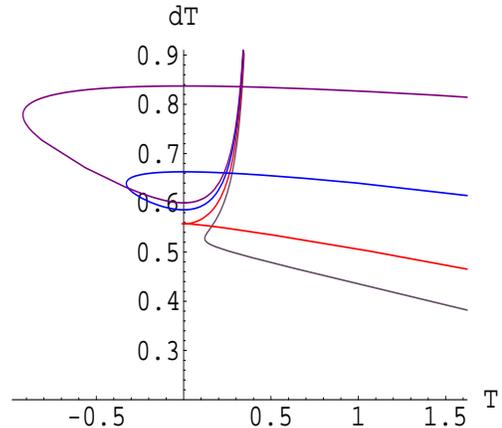,height=2.3in,width=3.0in}
\caption{($d=6$) The temperature differential vs temperature with
$\alpha_1 =\alpha_2 \equiv \alpha$. From top to bottom
$\alpha=0.12,\,0.16,\,0.21,\,0.25$} \label{figure19}
\end{figure}

\begin{figure}[ht]
\epsfig{figure=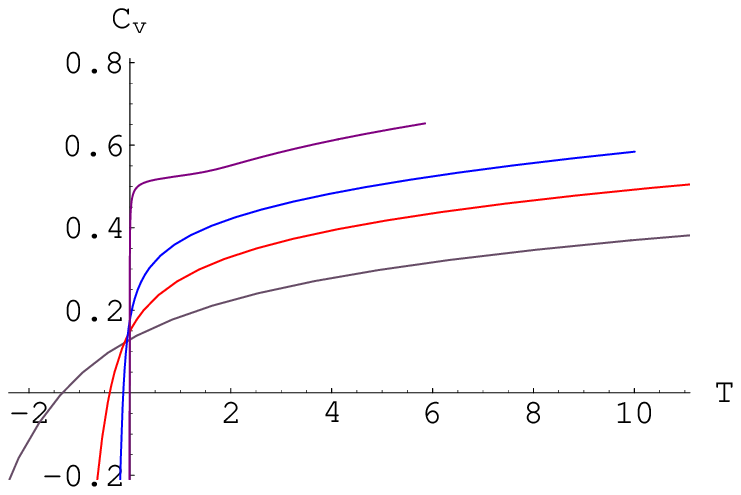,height=2.3in,width=3.0in}
\caption{($d=6$) The specific heat vs temperature with
$\alpha_1=\alpha_2=\alpha$. From top to bottom
$\alpha=0.25,\,0.5,\,0.6,\,0.7$} \label{figure20}
\end{figure}

\subsection{The specific heat and thermodynamic stability}

A black hole as a thermodynamic system is unstable if it has
negative specific heat. As it known, small Schwarzschild-AdS black
holes (i.e. with $a_i=0$) have negative specific heat but large
size black holes have positive specific heat. While there also
exists a discontinuity of the specific heat as a function of
temperature at $R=1/\sqrt{2}$, and so small and large black holes
are found to be somewhat disjoint objects. However, this is
essentially not the case when some of $a_i$ are non-trivial, and
especially, when $1\gtrsim \alpha_i\gg 0$, e.g., the small Kerr
black holes in $AdS_5$ space also have positive specific heat.

To this end, we shall study the thermodynamic stability of a
Kerr-AdS black hole by evaluating its specific heat, which is
given by
\begin{equation}
C_v=\frac{\partial E}{\partial T}.
\end{equation}
Figures~\ref{figure13} and \ref{figure14} show the plots of energy
and temperature differentials as functions of the horizon size
$R$. In the $d=5$ case, with equal rotation parameters, there is
clearly a minimum $R$, below which the temperature diverges. There
is also a minimum value of rotation parameter below which $dT$ can
be negative, which is $ \alpha\approx 0.17$ in five dimensions.
Above this value, both the temperature differential and the
specific heat are positive, see Figs.~\ref{figure15} and
\ref{figure16}. When plotted as a function of temperature, the
minimum of energy corresponds to the minimum in temperature, and
hence the specific heat is a monotonically increasing function of
Hawking temperature. This is also the case with a single rotation
parameter (see Fig.~\ref{figure17}), but now the critical value is
$\alpha\approx 0.25$.

It should also be noted that, with $d=5$ and $\alpha_i=\alpha \leq
1/6$, the thermodynamic behavior of a Kerr-AdS black hole at high
temperature can be very different from that at low temperature
(see Fig.~\ref{figure17}). A similar behavior is found in the case
of a single non-vanishing rotation parameter, though up to a
slightly larger value of $\alpha$ ($\lesssim 1/4$). Put another
way, in the $AdS_5$ background, small rotating black holes are
unstable only for rotation parameters of order $0.15\,l$ or less;
the precise limit is dimension dependent and black holes with
larger angular velocities are thermodynamically stable.

Similarly, in dimensions $d\geq 6$, the Kerr-AdS black holes
become unstable below some critical values of rotation parameters,
for which a new branch would appear. When $d=6$, with equal
rotation parameters, the critical value is $\alpha \approx 0.22$
(see Figs.~\ref{figure18}--\ref{figure20}) and it is slightly
higher in the case of a single rotation parameter.

Looking at the behavior of free energy and specific heat as
functions of horizon position $R$, we remark that a five
dimensional Kerr-AdS black hole with a single rotation parameter
is thermodynamically more stable over two (equal) rotation
parameter solutions. But this is essentially not the case in
dimensions six or more.

In all odd dimensions, the specific heat has a single branch at
high rotation but two branches at low rotations: the critical
value of $\alpha$ which distinguishes these two cases increases
with the number dimensions, and also with number of non-trivial
rotation parameters. A similar behavior is observed in all even
dimensions $d\geq 6$, but in this case an interesting difference
is that the specific heat can never be zero with $T>0$.

It seems relevant to ask what happens at the critical angular
velocity limit, $\alpha_i= 1$. Apparently, the action as well as
the entropy is divergent in this limit. Nevertheless, as discussed
in~\cite{Hawking98a} (see also \cite{Cvetic:2005nc}), there exists
a scaling of the mass parameter $m\to 0$ which makes the physical
charges of the configuration finite. With equal rotation
parameters, when $\alpha_i\to 1$, a Kerr-AdS black hole is more
preferred than a thermal AdS phase even at low temperature. In
fact, in all dimensions $d\geq 6$, small Kerr-AdS black holes with
a single non-vanishing rotation parameter are unstable.

In our plots we have used the energy expressions suggested by
Gibbons {\it et al.}~\cite{Gibbons04c}, which differ from those
suggested by Hawking {\it et al.}~\cite{Hawking98a} by some
overall constant factors. This itself does not introduce any
significant difference in the behavior of specific heat and hence
the thermodynamic stability of Kerr-AdS solutions. At any rate,
the energy measured in a non-rotating frame appeared more
suggestive to be used because it can be derived using various
other
methods~\cite{Henneaux85a,Ashtekar:1999jx,Mann00a,Deruelle04a,Deser:2005a};
the energy (or total mass) expressions given in~\cite{Mann00a},
however, disagree with those in~\cite{Gibbons04c} in odd spacetime
dimensions.

\subsection{The temperature bound for rotating black holes}

\begin{figure}[ht]
\epsfig{figure=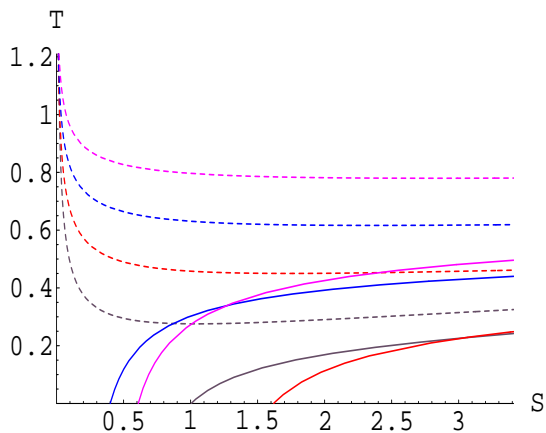,height=2.3in,width=3.0in}
\caption{The temperature vs entropy, in various dimensions with
equal rotation parameters, $\alpha_i\equiv \alpha$. From top to
bottom: $d=7,\,6,\,5,\,4$ with all $\alpha_i=0$ (dashed lines);
from left to right: $d=6,\,7,\, 4,\,5$ (solid lines) each with
$\alpha=0.4$.} \label{figure21}
\end{figure}

\begin{figure}[ht]
\epsfig{figure=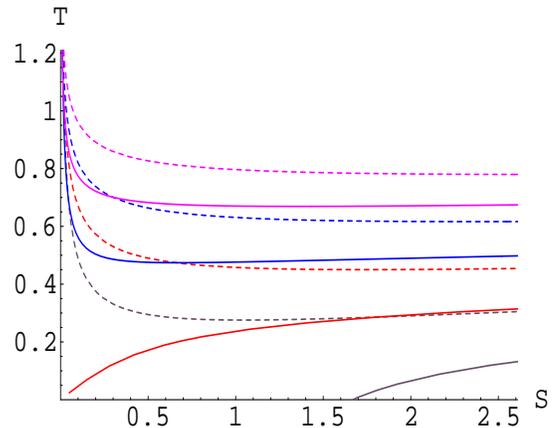,height=2.3in,width=3.0in}
\caption{The temperature vs entropy with a single rotation
parameter $\alpha$. From top to bottom: $d=7,\,6,\,5,\,4$ each
with $\alpha=0.4$ (solid lines), and $\alpha_i=0$ (dashed lines).}
\label{figure22}
\end{figure}

\begin{figure}[ht]
\epsfig{figure=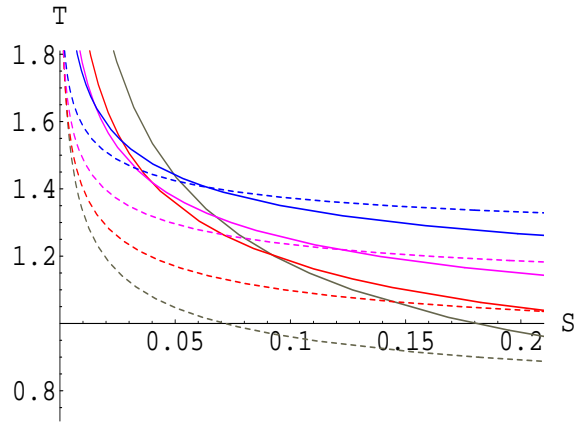,height=2.3in,width=3.0in}
\caption{The temperature vs entropy, in the small entropy region.
From top to bottom (in the region $S>0.1$): $d=10,\,9,\,8,\,7$
with $\alpha_1=\cdots \alpha_{N-1}=0.01$, $\alpha_N=0.9$ (solid
lines, Kerr-AdS) and $\alpha_i=0$ (dashed lines,
Schwarzschild-AdS).} \label{figure23}
\end{figure}

It was shown recently in \cite{Gibbons05a} that at fixed entropy,
the temperature of a rotating black hole is bounded above by that
of a non-rotating black hole, in four and five dimensions, but not
in six or more dimensions. We verify this claim by plotting
temperature as a function of entropy, in various dimensions; some
of the plots are depicted in Figs.~\ref{figure21}-\ref{figure23}.
In dimensions six or more, the minimum of entropy is not always
the minimum of temperature, it actually depends upon the choice of
rotation parameters. This is precisely the case where the
inequality $T_{\text{Kerr-AdS}} \geq T_{\text{S-AdS}}$ may be
realized with a very small entropy. But in this limit the
temperature actually diverges, so the effect like this might be
absent in a physical picture. At fixed entropy, but $S\gg 0$, the
Hawking temperature of a rotating black hole is always suppressed
relative to that of a non-rotating black hole and the inequality
$T_{\text{Kerr-AdS}} < T_{\text{S-AdS}}$ holds in all dimensions.
This result, presumably, holds with various charges and classical
matter fields (such as gauge fields, dilaton, etc) and is in
accord with the earlier observation made by Visser while studying
a static spherically symmetric case in four dimensions with no
cosmological term~\cite{Visser92a}.

A five dimensional Kerr-AdS black hole with a single non-vanishing
rotation parameter possesses an interesting (and perhaps
desirable) feature; in this case the entropy vanishes when the
temperature becomes zero. A similar feature is present in seven
dimensions, but with two equal rotation parameters:
$\alpha_1=\alpha_2\goesas 0.33$, $\alpha_3=0$.

In a recent work~\cite{K-i.Maeda05a}, on the evolution of a five
dimensional rotating black hole via the scalar field radiation,
Maeda~\etal ~observed that, in a flat background ($c=0$), the
asymptotic state of a five dimensional rotating black hole with a
single non-vanishing parameter is described by $a\goesas 0.11
\sqrt{M}$. It would be interesting to know a similar result in an
anti-de Sitter background, $c>0$.

\subsection{Rotation and the AdS-CFT correspondence}

Following~\cite{Witten98,Hawking98a}, one would expect the
partition function of a Kerr-AdS black hole to be related to the
partition function of a CFT in a rotating Einstein universe on the
(conformal) boundary of the AdS space.

A curious observation in Ref.~\cite{Cai:2005kw} is that the
Cardy-Verlinde entropy formula works more naturally using the bulk
thermodynamic variables defined by Hawking {\it et
al.}~\cite{Hawking98a}. This seems to indicate that the energy
expression (\ref{Energy-Hawking}) is still relevant in a dual CFT.
The Killing vector is then given by
\begin{equation}
\chi=\frac{\partial}{\partial \tau}+ \Omega_i
\frac{\partial}{\partial\phi_i},
\end{equation}
where $\phi_i$ are the angular coordinates. This property normally
allows the thermal radiation to rotate with black hole's angular
velocity all the way to conformal infinity.

One could ask whether or not the bulk thermodynamic variables
suggested by Gibbons {\it et al.}~\cite{Gibbons04c}, which were
measured with respect to a frame that is non-rotating at infinity,
can be mapped onto the boundary CFT variables by using the usual
scaling argument. This does not seem to be the case as long as the
CFT is assumed to be on a surface of large $R$ in Boyer-Linquist
coordinates. However, such a mapping might exist when the CFT is
assumed to be on a large spherical surface, that is one for which
the coordinate $y=\text{constant}$ at large $y$. That is to say,
it is possible that the set of bulk variables for Kerr-AdS black
holes given by Gibbons {\it et al.}\cite{Gibbons04c}, in some
(modified) form, match onto the boundary CFT variables that
satisfy the first law of thermodynamics. This was indeed shown to
be the case in~\cite{Gibbons05a}.

Let us briefly discuss the role of non-trivial rotation parameters
on the existence of an equilibrium between Kerr-AdS black hole and
rotating thermal radiation around it. For this the requirement of
a positive specific heat is a necessary condition. In five
dimensions, the specific heat is always positive and also a
monotonically increasing function of temperature when one (or
both) of the rotation parameters takes a value at least
one-quarter the AdS length scale $l$. This means, unlike in
Minkowski (infinite) space, the rotating Kerr-AdS black holes can
be in equilibrium with rotating thermal radiation around it, when
$0 \ll\alpha_i \lesssim 1 $; that is, the rotation parameter is
not significantly smaller than the AdS curvature radius, so as to
attain a stable equilibrium.

\medskip

\section{Stability of Kerr spacetime
under gravitational perturbations}

In this Section we study the gravitational stability of Kerr-AdS
background metrics (with $M=0$) in dimensions five and higher. For
this purpose, it is sufficient to consider the following
$d$-dimensional (time-independent) metric {\it Ansatz}:
\begin{equation} \label{metric-gen}
g_{a b}(X) dX^a dX^b=
g_{\mu\nu}(x) dx^\mu dx^\nu
+\gamma(x)^2 d\Sigma_{k,n}^2(\tilde{x}),
\end{equation}
where the metric $g_{a b}(X)$ is effectively separated into two
parts: a diagonal ``bulk'' line element and $d\Sigma_{k,n}^2$,
which is the metric on an $n$-dimensional base manifold whose
curvature has not been specified, (so $k=0$ or $\pm 1$), and hence
can be replaced by any Einstein-K\"ahler metric with the same
scalar curvature. However, in the present work we study only the
$k=+1$ case, and hence the base ${\cal M}^n$ may be viewed as an
$S^1$ fiber over $S^{n-1}$ (for odd $n$) or as $S^{n}$ (for even
$n$). For example, in the $d=5$ (or $n=3$) case, the event horizon
is $S^1\times S^2$.

Under a small linear metric perturbation
$$g_{ab}(X) \to g_{ab}(X)+h_{ab}(X) $$
with $|h_a^b|\ll 1$, the variation in the Ricci tensor is given by
\begin{equation}
\delta R_{ab}=\frac{1}{2}\Delta_L h_{ab}
-\frac{1}{2}\nabla_a\nabla_b h^c_c + \nabla_{(a}\nabla^c
h_{b)c},
\end{equation}
where the spin-2 Lichnerowicz operator $\Delta_L$ is defined by
(see, for example,~\cite{Ish:2001dv})
\begin{equation}
\Delta_L h_{ab} =
-\nabla^2 h_{ab} -2 R_{cadb} h^{cd}
+2 R_{c(a} h^c_{b)}.
\end{equation}
The stability of background metrics of the form~(\ref{metric-gen})
with $n>2$, under certain metric perturbations, is specific to
tensor perturbations. We therefore would like to restrict our
analysis here to the tensor mode fluctuations that satisfy
$$ h_{ab}(X)=0$$
unless $(a, b)= (i, j)$, where the indices $a, b, \cdots $ run
from $0 \cdots (d-1)$ and the indices $i, j, \cdots $ will run
from $(d-n) \cdots (d-1)$ for the $n$-dimensional base. The
variation of the Ricci tensor on the base ${\cal M}^{n}$ must then
satisfy
\begin{equation}\label{perturbed1}
\delta R_{i j}=\frac{1}{2} (\Delta_L  h)_{i j}=-c (d-1)
h_{i j},\end{equation}
where $c$ is the $d$-dimensional cosmological constant, with
\begin{eqnarray}
\Delta_L h_{ij}&=& \frac{1}{\gamma^2}\tilde{\Delta}_L h_{ij}
+\left[-g^{\mu\nu}\partial_\mu \partial_\nu\right]
h_{i j}\nonumber \\
&+& \sum_{\nu=1}^{d-n}\left[\partial^\sigma g_{\sigma \nu}
-\frac{1}{2}g^{\sigma\rho}\partial_\nu g_{\sigma\rho}
+(4-n)\frac{\partial_\nu\gamma}{\gamma}\right]\partial^\nu
h_{ij}\nonumber \\
&-& \frac{4}{\gamma^2}\left[g^{\mu\nu}\partial_\mu \gamma(x)
\partial_\nu \gamma(x)\right] h_{i j},
\end{eqnarray}
where $\tilde{\Delta}_L h_{i j}$ is the spin-2 Lichnerowicz
operator acting on the base ${\cal M}^{n}$. The Lichnerowicz
operator $\Delta_L $ is compatible with the transverse, trace-free
(de Donder) gauge for $h_{a b}$: $h_a^a=0=h^a_{b; a}$, see
e.g.~\cite{Gregory93a}.

\subsection{Dependence on radial coordinate only}

Let us first consider a background spacetime where $d=n+2$, such that
we can write the metric as
\begin{equation}
ds^2=-\alpha(r)^2 dt^2+\beta(r)^2 dr^2+\gamma(r)^2 d\Sigma_{n}^2.
\end{equation}
We can write the Lichnerowicz
operator as
\begin{eqnarray}
\Delta_L h_{i j} &=&
\frac{1}{\gamma^2} \tilde{\Delta}_L h_{i j}
+\left[\frac{\partial_t^2}{\alpha^2}
-\frac{\partial_r^2}{\beta^2}\right] h_{i j} \nonumber \\
&+& \left[\frac{\beta_r}{\beta}-\frac{\alpha_r}{\alpha}
+(4-n)\frac{\gamma_r}{\gamma}\right]
\frac{\partial_r h_{i j}}{\beta^2}\nonumber\\
&-& \frac{4}{\gamma^2} \frac{\gamma_r^2}{\beta^2}
\,h_{i j},
\end{eqnarray}
where the subscripts $t$, $r$ denote derivatives w.r.t. $t$, $r$
respectively. In this case we find it convenient to choose
\begin{equation}\label{trans1}
h_{i j}= \Psi(r) {\rm e}^{\omega t} \tilde{h}_{ij}(\tilde{x}),
\end{equation}
such that
\begin{equation}
(\tilde{\Delta}_L \tilde{h})_{ij} =\lambda \tilde{h}_{ij},
\end{equation}
where $\tilde{x}$ are coordinates on ${\cal M}^{n}$ and
$\lambda$ is the eigenvalue of the Lichnerowicz operator on
${\cal M}^n$. We want
to write the perturbed equations~(\ref{perturbed1}) in the form:
\begin{equation}
\left(\partial_{r_*}^2-V(r_*)\right)\Phi(r_*) = \omega^2\Phi(r_*).
\end{equation}
To facilitate this we introduce two transformations:
\begin{equation}\label{tranf2}
dr=\frac{\partial r}{\partial r_*} dr_*,\quad \Psi(r)=\chi(r)
\Phi(r)
\end{equation}
with
\begin{equation}
\chi(r)= C_1 {\gamma}^{(4-n)/2}.
\end{equation}
We then find (see Appendix for details)
\begin{eqnarray}
V(r(r_*))&=& \frac{\lambda \alpha^2}{\gamma^2} +\frac{n^2-10 n+8}{4}
\left(\frac{\gamma_{r_*}}{\gamma}\right)^2\nonumber \\
&{}& +\,\frac{(n-4)}{2} \frac{\gamma_{r_* r_*}}{\gamma} + 2(n+1)c \alpha^2,
\end{eqnarray}
where,
\begin{eqnarray}
\gamma_{r_*} &\equiv& \frac{\partial{r}}{\partial{r_*}} \frac{\partial
\gamma}{\partial r}
=\frac{\alpha}{\beta} \gamma_r, \nonumber \\
\gamma_{r_* r_*} &=& \frac{\alpha^2}{\beta^2}\left[\gamma_{r
r}+\left(\frac{\alpha_r}{\alpha}-\frac{\beta_r}{\beta} \right)
\gamma_r\right].
\end{eqnarray}
The above potential correctly reproduces the result
in~\cite{Ish03a} (cf. Eq.~(41) with $\alpha^2=f(r)$ and
$\gamma^2=r^2$), see also~\cite{Hartnoll02a,Kodama:2003jz}.
Apparently, the case $n=4$ is special.

\subsection{Kerr-AdS backgrounds in odd dimensions}

While we believe the stability analysis of Kerr-AdS background
metrics can be generalized to non-equal rotation parameters (or
angular momenta), we shall focus on the case with equal rotation
parameters.

In the case of an odd number of spacetime dimensions $d=2N+1=n+2$,
the zero-mass ($M=0$) Kerr (Anti)-de Sitter background metric may
be given by
\begin{eqnarray}
ds^2 &=& -\frac{(1+cr^2)\, dt^2}{(1-ca^2)} +
\frac{r^2\,dr^2}{(1+cr^2)(r^2+a^2)} \nonumber \\
&{}& +\,\frac{r^2+a^2}{1-ca^2} ds^2({\cal M}^{n}),
\end{eqnarray}
where the rotation parameters are set equal (i.e., $a_1=a_2=a$).
The base space ${\cal M}^{n}$, which is topologically $S^{2N-1}$,
may be parameterized by the metric
\begin{equation}
(d\psi+A)^2+d\Sigma_{N-1}^2
\end{equation}
where $d\Sigma_{N-1}^2$ is the canonically normalized Fubini-Study
metric on an $(N-1)$ dimensional complex projective space
$\CI\PI^{N-1}$, and $A$ is a local potential for the K\"ahler form
$J=\frac{1}{2} dA$ on $\CI\PI^{N-1}$. For example, in five
dimensions, the metric on base ${\cal M}^3$ is $ds^2 ({\cal M}^3)=
d\theta^2+\sin^2\theta d\phi^2 +\cos^2\theta d\psi^2$.

In the above background, the linear tensor perturbations satisfy
\begin{equation}\label{Lich-eq}
\Delta_L h_{i j} =- 2 c\,(n+1)\, h_{i j},
\end{equation}
where
\begin{eqnarray}
\Delta_L h_{i j}&=&
\bigg[-\frac{(r^2+a^2)(1+cr^2)}{r^2}
\left(\frac{\partial^2}{\partial{r}^2}+\frac{4
r^2}{(r^2+a^2)^2}\right)
\nonumber \\
&{}& - \left((n-2)c r+ \frac{n-4}{r}- \frac{a^2(1-cr^2)}{r^3}\right)
\frac{\partial}{\partial{r}}\nonumber \\
&{}&+\, \frac{1-ca^2}{1+cr^2} \frac{\partial^2}{\partial{t}^2}\bigg]
h_{i j} + \frac{1-ca^2}{r^2+a^2} (\tilde{\Delta}_L h)_{i j}.
\end{eqnarray}
In terms of the Regge-Wheeler type coordinate $r_*$, which may be
defined by
\begin{equation}
{dr}=\frac{(1+cr^2)\sqrt{r^2+a^2}}{r \sqrt{1-ca^2}}\, {dr_*},
\end{equation}
and using Eqs.~(\ref{trans1}),(\ref{tranf2}) the differential
equation is cast in the standard form:
\begin{equation}\label{schro r}
-\frac{d^2\Phi}{d r_*^2} + V(r(r_*))\Phi=- \omega^2 \Phi \equiv
E^2\Phi,
\end{equation}
where the potential is
\begin{eqnarray}\label{potr1}
 V(r(r_*)) &=& \frac{\lambda(1+c r^2)}{r^2+a^2}
+\frac{(n^2- 10 n+8)(1+c r^2)^2}{4(1-c a^2)(r^2+a^2)}\nonumber \\
 &{}&~~~~~~~~+\,\frac{(3 n-2) c (1+c r^2)}{1-c a^2}.
\end{eqnarray}
This potential is well behaved around $r=0$ unlike for the
AdS-Schwarzschild metric (i.e. $a=0$).

There exists a criterion for stability (e.g. the Schr\"odinger
equation not possessing a bound state with $\omega>0$), in terms
of the minimum Lichnerowicz eigenvalue, $\lambda_{min}$, on the
base manifold ${\cal M}^{n}$. In the case of a vanishing
cosmological constant ($c=0$), this criterion is the same as that
for a Schwarzschild-AdS background~\cite{Hartnoll02a}:
 \begin{equation}\label{instab}
\lambda_{min}\geq\lambda_c =4-\frac{(5-n)^2}{4}\Leftrightarrow
\textrm{stability}.
\end{equation}
A requirement that $\lambda_c \geq 0$ constrains the spacetime
dimensions to $n\leq 9$ (or $d\leq 11$). The stability of a
potential depends on the eigenvalue $\lambda$, ensuring that the
potential is positive everywhere and bounded from below. Defining
$\mu\equiv c a^2$, with $a>0$, we require $\mu<1$ for $c>0$ and we
find
\begin{equation}
\lambda \geq \tilde{\lambda}_c=-\frac{n^2-10
n+8+4(3n-2)\,\mu}{4(1-\mu)}. \label{lambdaminr}
\end{equation}
The lower bound on $c a^2$ required for gravitational stability of
the background metric is found to be stronger than that for
thermodynamic stability. In the de Sitter case (i.e. $c<0$), there
is a mass gap, so $\lambda$ starts from a finite value, and $c
a^2$ is unbounded from below.

Instead of solving the Schr\"odinger equation directly in terms of
$r_*$, one can solve the radial part of equation~(\ref{Lich-eq})
by expressing it as a hypergeometric equation, whose solution is
given by a linear combinations of
\begin{eqnarray}
\Psi_{\pm}(x,\mu) &=& \left(\frac{x+\mu}{c}\right)^{(5-n\pm
2\nu)/4}
\left(1+\mu\right)^{i\,\omega/2\sqrt{c}}\nonumber \\
&{}& {}_2F_1\bigg(\frac{\pm
2\nu-(n-1)}{4}+\frac{i\,\omega}{2\sqrt{c}},\nonumber \\
&{}& \frac{\pm 2\nu+(n+3)}{4}+\frac{i\,\omega}{2\sqrt{c}},\pm
\nu+1;
-\frac{x+\mu}{1-\mu}\bigg),\nonumber\\
\end{eqnarray}
where $x\equiv c r^2$, and
\begin{equation}
\nu=\frac{1}{2}\sqrt{4\lambda+(5-n)^2-16}.
\end{equation}
We note that reality of $\nu$ immediately gives the stability
condition~(\ref{instab}). Reality of the solution also requires
$\omega=i\tilde{\omega}$ which implies that there are no
exponentially growing (unstable) modes. Requiring the solution to
be bounded as $r\rightarrow \infty$ fixes one arbitrary constant
which leaves $\Psi$ decaying as $r^{1-n}$. Given that $\Psi=\chi
\Phi$ we find that $\Phi$ decays as $r^{-(n+2)/2}$. By considering
the large $r$ limit of potential~(\ref{potr1}) we also see that $n
\geq 2$ so that Eq~(\ref{schro r}) remains bounded as required to
make the total energy finite.

\medskip
\subsection{Kerr-AdS backgrounds in even dimensions}

Consider a background spacetime where $d=n+3$, such that
we can write the metric as
\begin{eqnarray}
ds^2&=& -\alpha(r,\theta)^2 dt^2+\beta(r,\theta)^2 dr^2+
\sigma(r,\theta)^2 d\theta^2\nonumber \\
&{}& +\gamma(r,\theta)^2 d\Sigma_{n}^2.
\end{eqnarray}
We can write the Lichnerowicz
operator as
\begin{eqnarray}\label{mainLich1}
\Delta_L h_{i j}&=&\frac{1}{\gamma^2} \tilde{\Delta}_L h_{i j}
+\left[\frac{\partial_t^2}{\alpha^2}-\frac{\partial_r^2}{\beta^2}-
\frac{\partial_\theta^2}{\sigma^2}\right]h_{i j}~~~~~~~~\nonumber \\
&+&\left[-\frac{\alpha_r}{\alpha}+\frac{\beta_r}{\beta}
-\frac{\sigma_r}{\sigma}
+(4-n)\frac{\gamma_r}{\gamma}\right] \frac{{\partial_r} h_{i
j}}{\beta^2}\nonumber
\\
&+&\left[-\frac{\alpha_\theta}{\alpha}
-\frac{\beta_\theta}{\beta}+\frac{\sigma_\theta}{\sigma}
+(4-n)\frac{\gamma_\theta}{\gamma}\right]\frac{\partial_\theta h_{i
j}}{\sigma^2}\nonumber
\\
&-&\frac{4}{\gamma^2}
\left[\frac{\gamma_r^2}{\beta^2}+\frac{\gamma_\theta^2}{\sigma^2}\right]
h_{i j},
\end{eqnarray}
To this end, we shall consider a Kerr-AdS background metric
($M=0$) in even dimensions, $n= 2N-1$, by setting the $N$ rotation
parameters equal (i.e. $a_1=\cdots =a_{N}=a$). The background
metric is~\cite{Gibbons04a}
\begin{eqnarray}\label{background-d-even}
ds^2&=&-\frac{(1+cr^2)\Delta_\theta}{1-ca^2}
{dt}^2+\frac{\rho^2}{(1+cr^2)(r^2+a^2)}{dr}^2\nonumber \\
&{}&+ \frac{\rho^2}{\Delta_\theta} {d\theta}^2
+\frac{(r^2+a^2)\sin^2\theta}{1-ca^2} \overline{ds}^2({\cal
M}^{n}),
\end{eqnarray}
where,
\begin{equation}
\rho^2\equiv r^2+a^2 \cos^2\theta, \quad \Delta_\theta=1-ca^2
\cos^2\theta.
\end{equation}
A calculation gives
 \begin{eqnarray}\label{main-eq-r-theta}
\left(\Delta_L h\right)_{i j}&=& \bigg[ \frac{1-ca^2}{(1+ c
r^2)\Delta_\theta} \frac{\partial^2}{\partial{t}^2}-
\frac{(1+cr^2)(r^2+a^2)}{\rho^2}
\frac{\partial^2}{\partial{r}^2}\nonumber\\
&-& \frac{\Delta_\theta}{\rho^2}
\frac{\partial^2}{\partial{\theta}^2} -\frac{4}{\rho^2}
\left(\frac{r^2(1+c r^2)}{r^2+a^2}+\frac{\Delta_\theta}{\tan^2\theta}
\right)\bigg] h_{i j}\nonumber \\
&+& \frac{(1-c a^2)}{(r^2+a^2)\sin^2\theta}
\tilde\Delta_L h_{i j}\nonumber \\
&+& \frac{r}{\rho^2} \left( 2(1-c a^2)-(n-1)(1+c r^2)\right)
\partial_r h_{i j}\nonumber \\
&+&\frac{1}{\rho^2\, \tan\theta}\left(2(1-c a^2)-(n-2)\Delta_\theta
\right)\partial_\theta h_{i j}.\nonumber \\
\end{eqnarray}
Equation~(\ref{main-eq-r-theta}) may be separated by writing,
\begin{equation}
h_{i j}= \Psi(r) {\rm e}^{\omega t} S(\theta)
\tilde{h}_{ij}(\tilde{x}),
\end{equation}
and taking the large $r$ limit. Hence,
\begin{eqnarray}
&& \left(r^2 \frac{\partial^2}{\partial r^2} +(n-1)r
\frac{\partial}{\partial r}-2n
+\frac{p}{c r^2}\right)\Psi=0,\\
&&
\Delta_\theta \frac{\partial^2 S}{\partial\theta^2}
-\frac{1}{\tan\theta}\left(2(1-c a^2)-(n-2)\Delta_\theta\right)
\frac{\partial S}{\partial\theta} \nonumber \\
&& +\left(\frac{4\Delta_\theta}{\tan^2\theta}-\frac{\lambda (1-c
a^2)}{\sin^2\theta}  -\frac{\omega^2 (1-c
a^2)}{c\Delta_\theta}- p\right)S=0, \label{angular-eq}\nonumber \\
\end{eqnarray}
where we have defined $(\tilde{\Delta}_L \tilde{h})_{ij} =\lambda
\tilde{h}_{ij}$, so that $\lambda$ is the eigenvalue of the
Lichnerowicz operator on ${\cal M}^n$, and $p$ is the separation
constant.

The radial equation is easily solved to yield
\begin{eqnarray}\label{radial-sol}
\Psi &=& c_1\, r^{(2-n)/2} J_1
\left(\frac{n+2}{2},\sqrt{\frac{p}{c r^2}}\right)\nonumber
\\
&{}&+\, c_2\, r^{(2-n)/2} \, Y_1
\left(\frac{n+2}{2},\sqrt{\frac{p}{c r^2}}\right).
\end{eqnarray}
However, regularity of the radial solution at $r=\infty$ requires
$c_2=0$ and hence as $r\to \infty$ the radial solution behaves as
\begin{equation}
\Psi(r) \sim \frac{c_1}{r^{n}}.
\end{equation}
Equation~(\ref{angular-eq}), together with boundary conditions of
regularity at $\theta=0$ and $\pi$, constitute an eigenvalue
problem for the separation constant $p$. For $\sin\theta\approx
\theta, ~\cos\theta\approx 1$, the solution is
\begin{equation}
S = \theta^{(5-n)/2}\left[ c_1 J_m(z)+c_2
Y_m(z)\right],
\end{equation}
where
\begin{eqnarray}
m&=&\sqrt{\lambda-4+\frac{(5-n)^2}{4}},\nonumber \\
z&=& \sqrt{-\frac{p c +\omega^2}{c(1-c a^2)}}\,\theta.
\end{eqnarray}
The criterion for gravitational stability, in terms of the minimum
Lichnerowicz eigenvalue $\lambda_{min}$ on the base manifold
${\cal M}^n$, namely $\lambda_{min} \geq \lambda_c=4-(5-n)^2/4$,
now translates into the requirement that $m\in\mathbf{R}$. However
we note that $c\neq0$ in this case. In AdS space, since $c>0$, for
reality of the solution we also require,
\begin{equation}
0< 1- ca^2 < 1, \quad p < -\frac{\omega^2}{c}.
\end{equation}
For real $\omega$, $p<0$ and hence $\sqrt{{p}/{c r^2}}$ is
imaginary, but this is not allowed by the radial wave equation.
Therefore there are no normalisable solutions with
$\omega\in\mathbf{R}$. For $\omega\to i\tilde{\omega}$, one
requires $p<\tilde{\omega}^2/c$. A useful inequality for stability
of the background AdS metric~(\ref{background-d-even}) is
therefore,
\begin{equation}
0 < p<\frac{\tilde{\omega}^2}{c}.
\end{equation}

Instead of considering the large $r$ limit
in~(\ref{main-eq-r-theta}), let us now consider the special case where
the angular velocity approaches the critical limit, $c a^2=1$ (or
$a=l$). The eigenfunctions are then the associated Legendre
polynomials $P_{\tilde{n}}^m(\cos\theta)$,
$Q_{\tilde{n}}^m(\cos\theta)$, where,
\begin{eqnarray}
m&=&\frac{1}{2}\sqrt{4p-(7-n)(n+1)},\nonumber \\
\tilde{n}&=&\frac{1}{2}\left(\sqrt{(n-6)(n+2)}-1\right).
\end{eqnarray} An interesting case is $n=7$, which allows one to
study supergravity solutions in $d=10$. It would be interesting to
know what the limit $c a^2\to 1$ corresponds to in a dual field
theory. We leave this issue to future work.

\section{Conclusion}

In this paper we have studied the thermodynamics and stability of
higher-dimensional ($d\geq 5$) rotating black holes in a
background (anti)-de Sitter spacetime. The thermodynamic
quantities for Kerr-AdS black hole solutions suggested by Gibbons
{\it et al.} ~\cite{Gibbons04c} have been used to study the
behavior of the free energy and specific heat (which are defined
unambiguously in all spacetime dimensions $d\geq 4$) as functions
of temperature and horizon positions. The two apparently different
expressions of energy in the Kerr-AdS background suggested by
Hawking {\it et al.}~\cite{Hawking98a} and Gibbons {\it et
al.}~\cite{Gibbons04c} do not introduce any significant difference
in the behavior of bulk thermodynamic quantities (such as entropy,
free energy, specific heat, etc) and therefore the stability of
Kerr-AdS solutions. Nevertheless, the Gibbons {\it et al.} bulk
variables are more suggestive to be used as they map onto the
boundary variables with the natural definition of conformal
boundary metric, that is the one for which the coordinate
$y=\text{constant}$ for large $y$, and satisfy the first law of
thermodynamics.

As for thermodynamic stability, rotating black holes are found to
be stable down to a critical value of the rotation parameter,
below which the specific heat becomes negative. For example, a
five dimensional Kerr-AdS black hole is thermodynamically stable
when the rotation parameters take values $a_i \approx 0.17\,l$;
larger angular velocities usually stabilize the black hole.

Similarly, a zero-mass Kerr-AdS background is gravitationally
stable down to a critical eigenvalue, below which the
Schr\"odinger equation may involve growing tensor mode
perturbations. Again larger angular velocities stabilize the
background Kerr-AdS spacetimes, although the bound on the rotation
parameters required for the gravitational stability of rotating
black holes is not directly related to that of thermodynamic
stability.

An obvious extension to study, for completeness, would be the
inclusion of non-zero charges and all of the possible rotation
parameters in dimensions five and higher. A particularly
interesting problem would be a study of the gravitational
stability of \textit{massive} Kerr-AdS black hole spacetimes in
$AdS_5$ and $AdS_7$ spaces. Some of the problems will be discussed
in a follow-up paper~\cite{IshBen05b}.

\bigskip

{\bf Acknowledgements}\ ~It is a pleasure to thank Roy Kerr and
David Wiltshire for discussions. I.P.N. also wishes to thank Gary
Gibbons for useful discussions and Chris Pope for helpful remarks.
This work was supported in part by the Mardsen fund of the Royal
Society of New Zealand.

\renewcommand{\theequation}{A.\arabic{equation}}\setcounter{equation}{0}

\section*{Appendix:~Schr\"odinger Equation}

Consider a second order differential equation of the form
\begin{equation}
(A \partial_r^2 + B \partial_r + C + D
\partial_t^2+E \tilde\Delta_L )h =0,
\end{equation}
where $A,\,B,\,C$ are functions of $r$ only. We find it is
convenient to choose $h\equiv \Psi(r) {\rm e}^{\omega t}
\tilde{h}$, such that $\tilde\Delta_L \tilde{h}=\lambda\tilde{h}$.
We then have
\begin{equation}
(A \partial_r^2 +B \partial_r + C+ D \omega^2+ E \lambda )\Psi(r)
{\rm e}^{\omega t} \tilde{h} =0.
\end{equation}
\\
For non-zero fluctuations, ${\rm e}^{\omega t} \tilde{h}\neq
0$, this implies that
\begin{equation}
(A \partial_r^2 + B \partial_r + \tilde{C} )\Psi(r) =0,
\end{equation}
where $\tilde{C}\equiv C+ D \omega^2+ E \lambda$. We would like to
write this in the form
\begin{equation}
\left(\partial_{x}^2 - V(x(r))\right)\varphi =\omega^2 \varphi.
\end{equation}
To facilitate this we introduce two transformations:
$$ dr=\frac{\partial r}{\partial x} dx, \quad
\Psi=\chi\varphi.$$
The differential equation then takes the form
\begin{eqnarray}
&&\frac{A}{r_x^2} \varphi^{\prime\prime} +\left[\frac{2
A}{r_x^2}\frac{\chi^\prime}{\chi} +\frac{B}{r_x}
-\frac{A\,r_{xx}}{r_x^3}\right]
\varphi^\prime \nonumber \\
&&~~+\left[\frac{A}{r_x^2}\frac{\chi^{\prime\prime}}
{\chi}+\left(\frac{B}{r_x}-\frac{A\,r_{xx}}
{r_x^3}\right)\frac{\chi^\prime}{\chi}+
\tilde{C}\right]\varphi=0,\nonumber \\
\end{eqnarray}
where $r_x\equiv r^{\prime}= (\partial r/\partial x)$. Let us define
$$
r_x^2=-\frac{A}{D}, \quad \frac{\chi^\prime}{\chi}=\frac{B}{2 D
r_x}+\frac{r_{xx}}{2
  r_x}.$$
This implies
\begin{equation}
\frac{r_{x x}}{r_x}=\frac{1}{2}
\left(\frac{A_x}{A}+\frac{D_x}{D}\right).
\end{equation}
The differential equation then takes the standard form:
\begin{equation}
\partial_x^2\varphi - V\varphi=\omega^2 \varphi,
\end{equation}
where
\begin{equation}
V=-\left(\frac{\chi^{\prime}}{\chi}\right)^\prime
+\left(\frac{\chi^\prime}{\chi}\right)^2 +\frac{\bar{C}}{D},
\end{equation}
where $\bar{C}\equiv C+ E \lambda$.


\end{document}